\documentclass[twocolumn,preprintnumbers,prb,aps,amssymb,showpacs,superscriptaddress, altaffillsymbol]{revtex4-1}

\usepackage{graphicx} 
\usepackage{hyperref}
\usepackage{bm} 

\newcommand{\lsco}{La$_{2-x}$Sr$_x$CuO$_4$}

\newcommand{\pcco}{Pr$_{2-x}$Ce$_{x}$CuO$_{4}$}
\newcommand{\ybco}{YBa$_{2}$Cu$_{3}$O$_{y}$}
\newcommand{\tltwotwoone}{Tl$_{2}$Ba$_{2}$CuO$_{6+\delta}$}
\newcommand{\hbco}{HgBa$_{2}$CuO$_{4+\delta}$}
\newcommand{\bsco}{Bi$_2$Sr$_{2-x}$La$_x$CuO$_{6+\delta}$}

\begin{document}



\title{Wiedemann-Franz law in the underdoped cuprate superconductor \ybco}




\author{G.~Grissonnanche}
\affiliation{D\'{e}partement de physique  \&  RQMP, Universit\'{e} de Sherbrooke, Sherbrooke,  Qu\'{e}bec J1K 2R1, Canada}

\author{F.~Lalibert\'{e}}
\affiliation{D\'{e}partement de physique  \&  RQMP, Universit\'{e} de Sherbrooke, Sherbrooke,  Qu\'{e}bec J1K 2R1, Canada}

\author{S.~Dufour-Beaus\'{e}jour}
\affiliation{D\'{e}partement de physique  \&  RQMP, Universit\'{e} de Sherbrooke, Sherbrooke,  Qu\'{e}bec J1K 2R1, Canada}

\author{M.~Matusiak}
\affiliation{D\'{e}partement de physique  \&  RQMP, Universit\'{e} de Sherbrooke, Sherbrooke,  Qu\'{e}bec J1K 2R1, Canada}
\affiliation{Institute of Low Temperature and Structure Research, Polish Academy of Sciences, Wroclaw 50-950, Poland}

\author{S.~Badoux}
\affiliation{D\'{e}partement de physique  \&  RQMP, Universit\'{e} de Sherbrooke, Sherbrooke,  Qu\'{e}bec J1K 2R1, Canada}

\author{F.~F.~Tafti}
\affiliation{D\'{e}partement de physique  \&  RQMP, Universit\'{e} de Sherbrooke, Sherbrooke,  Qu\'{e}bec J1K 2R1, Canada}

\author{B.~Michon}
\affiliation{D\'{e}partement de physique  \&  RQMP, Universit\'{e} de Sherbrooke, Sherbrooke,  Qu\'{e}bec J1K 2R1, Canada}

\author{A.~Riopel}
\affiliation{D\'{e}partement de physique  \&  RQMP, Universit\'{e} de Sherbrooke, Sherbrooke,  Qu\'{e}bec J1K 2R1, Canada}

\author{O.~Cyr-Choini\`ere}
\affiliation{D\'{e}partement de physique  \&  RQMP, Universit\'{e} de Sherbrooke, Sherbrooke,  Qu\'{e}bec J1K 2R1, Canada}

\author{J.~C.~Baglo}
\affiliation{Department of Physics and Astronomy, University of British Columbia, Vancouver, British Columbia V6T 1Z4, Canada}

\author{B.~J.~Ramshaw}
\affiliation{Department of Physics and Astronomy, University of British Columbia, Vancouver, British Columbia V6T 1Z4, Canada}
\altaffiliation[Present address]{Los Alamos National Laboratory, Los Alamos, New Mexico 87545, USA.}

\author{R. Liang}
\affiliation{Department of Physics and Astronomy, University of British Columbia, Vancouver, British Columbia V6T 1Z4, Canada}
\affiliation{Canadian Institute for Advanced Research, Toronto, Ontario M5G 1Z8, Canada}

\author{D.~A.~Bonn}
\affiliation{Department of Physics and Astronomy, University of British Columbia, Vancouver, British Columbia V6T 1Z4, Canada}
\affiliation{Canadian Institute for Advanced Research, Toronto, Ontario M5G 1Z8, Canada}

\author{W.~N.~Hardy}
\affiliation{Department of Physics and Astronomy, University of British Columbia, Vancouver, British Columbia V6T 1Z4, Canada}
\affiliation{Canadian Institute for Advanced Research, Toronto, Ontario M5G 1Z8, Canada}

\author{S.~Kr\"{a}mer}
\affiliation{Laboratoire National des Champs Magn\'etiques Intenses, UPR 3228, (CNRS-INSA-UJF-UPS), Grenoble 38042, France}

\author{D.~LeBoeuf}
\affiliation{Laboratoire National des Champs Magn\'etiques Intenses, UPR 3228, (CNRS-INSA-UJF-UPS), Grenoble 38042, France}

\author{D.~Graf}
\affiliation{National High Magnetic Field Laboratory, Tallahassee, FL 32310, USA}

\author{N.~Doiron-Leyraud}
\affiliation{D\'{e}partement de physique  \&  RQMP, Universit\'{e} de Sherbrooke, Sherbrooke,  Qu\'{e}bec J1K 2R1, Canada}

\author{Louis~Taillefer}
\email{louis.taillefer@usherbrooke.ca}
\affiliation{D\'{e}partement de physique  \&  RQMP, Universit\'{e} de Sherbrooke, Sherbrooke,  Qu\'{e}bec J1K 2R1, Canada}
\affiliation{Canadian Institute for Advanced Research, Toronto, Ontario M5G 1Z8, Canada}


\date{\today}


\begin{abstract}
The electrical and thermal Hall conductivities of the cuprate superconductor YBa$_2$Cu$_3$O$_y$, $\sigma_{xy}$ and $\kappa_{xy}$,
were measured in a magnetic field up to
35~T, at a hole concentration (doping) $p = 0.11$.
In the $T = 0$ limit, we find that the Wiedemann-Franz law, $\kappa_{xy} / T = (\pi^2/3) (k_{\rm B} / e)^2 \sigma_{xy}$,
is satisfied for fields immediately above the vortex-melting field $H_{\rm vs}$.
This rules out the existence of a vortex liquid at $T = 0$ and it puts a clear constraint on the nature of the normal state in underdoped cuprates,
in a region of the doping phase diagram where charge-density-wave order is known to exist.
As the temperature is raised, the Lorenz ratio, $L_{\rm xy} = \kappa_{\rm xy} / (\sigma_{\rm xy} T)$, decreases rapidly, indicating that strong small-{\bf $q$} scattering processes are involved.
%
%
\end{abstract}







\maketitle



\section{INTRODUCTION}

The observation of a small electron pocket in the Fermi surface of underdoped \ybco~(YBCO)\cite{Leyraud2007Quantuma,Leboeuf2007Electron}
and \hbco~(Hg1201),\cite{doiron-leyraud_hall_2013,Barisic2013Universala}
in sharp contrast with the large hole-like Fermi surface of the overdoped regime,\cite{Vignolle2008Quantum}
shows that  the Fermi surface of hole-doped cuprates undergoes a profound transformation with underdoping.\cite{taillefer_fermi_2009}
In the cuprate La$_{1.8-x}$Eu$_{0.2}$Sr$_x$CuO$_4$,
the similar Fermi-surface reconstruction (FSR)\cite{cyr_choiniere_enhancement_2009,taillefer_fermi_2009,Chang2010Nernst,Laliberte2011FermiSurface}
is clearly linked to the onset of charge-stripe order detected by {X-ray} diffraction.\cite{cyr_choiniere_enhancement_2009,fink_phase_2011}

In YBCO, the recent detection of charge density-wave (CDW) modulations\cite{Wu2011MagneticFieldInduced,Ghiringhelli2012LongRangea,Chang2012Directa,Achkar2012Distinct} in the same doping range where the electron pocket prevails\cite{Leboeuf2011Lifshitz} points here also to a scenario where CDW order causes the FSR.
Moreover, recent evidence for an additional small hole-like pocket in the Fermi surface of YBCO (ref.~\onlinecite{doiron-leyraud_evidence_2015}) is consistent with calculations of FSR by the observed CDW order.\cite{Allais2014Connectinga}
Note that CDW order competes with superconductivity,\cite{Ghiringhelli2012LongRangea,Chang2012Directa,Achkar2012Distinct}
causing a suppression of the latter that is directly visible in the upper critical field $H_{\rm c2}$ measured as a function of doping,\cite{Grissonnanche2014Directa}
which exhibits a local minimum where the CDW is strongest.\cite{hucker_competing_2014,blanco-canosa_resonant_2014,Wu2013Emergence}

CDW modulations have also been seen in Hg1201, \cite{Tabis2014Charge}
\bsco~(Bi-2201),\cite{Comin2014Charge}
and \lsco~(LSCO),\cite{croft_charge_2014}
clear evidence that they are universal to hole-doped cuprates. This naturally begs the following question: what is the nature of the normal state of underdoped cuprates, in particular at low temperature, when superconductivity is suppressed by a magnetic field? Is it a dual state where charge order and superconductivity are intertwined, as proposed by certain theories\cite{Chen2004Pair,berg_dynamical_2007,Lee2014Amperean,Allais2014Connectinga}, or is it a metal without any superconducting component? More generally, does this metal depart from standard Fermi-liquid behaviour?

\begin{figure}[t]
\centering
\includegraphics[width=0.48\textwidth]{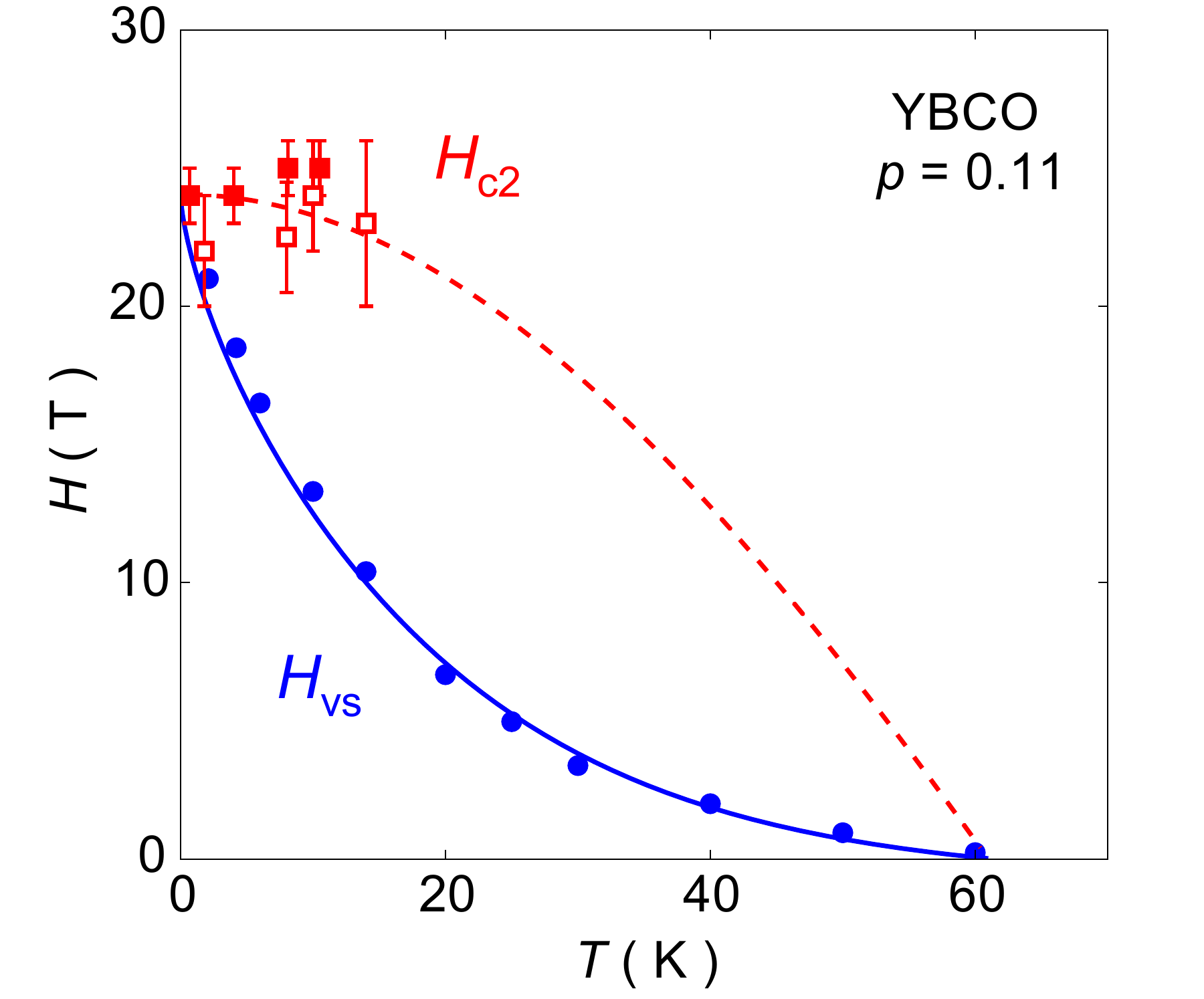}
\caption{
Magnetic field-temperature phase diagram of YBCO at a hole concentration (doping) $p = 0.11$,
showing the upper critical field $H_{\rm c2}(T)$ (red squares), detected in the longitudinal thermal conductivity $\kappa_{\rm xx}$
(open squares, from ref.~\onlinecite{Grissonnanche2014Directa}) and in the thermal Hall conductivity $\kappa_{\rm xy}$
(full squares, this work).
%
The red dashed line is a guide to the eye, showing how $H_{\rm c2}(T)$ might extrapolate to zero at $T_{\rm c}$.\cite{Grissonnanche2014Directa}
The blue symbols mark $H_{\rm vs}(T)$, the field at which the vortex solid melts and above which the electrical resistance is no longer zero;
the solid line is a fit to the theory of vortex-lattice melting.\cite{Ramshaw2012Vortex}
In the limit of $ T = 0$, the fit extrapolates to $H_{\rm vs}(0) = 24 \pm 2$~T, so that $H_{\rm vs}(T) = H_{\rm c2}(T)$ at $T = 0$.
}
\label{fig_1}
\end{figure}

The sharp suppression of the longitudinal thermal conductivity $\kappa_{\rm xx}$ with decreasing magnetic field recently observed in YBCO has been attributed to the onset  of vortex scattering, making it a direct measure of $H_{\rm c2}$.\cite{Grissonnanche2014Directa}
As shown in Fig. \ref{fig_1}, $H_{\rm c2}$ and the vortex-melting field $H_{\rm vs}$ were found to be equal at $T \rightarrow 0$, consistent with the absence of a vortex liquid at $T = 0$. Yet specific heat\cite{Riggs2011Heat} and magnetization\cite{Yu2014Diamagnetic} data have been interpreted in terms of superconductivity persisting well above $H_{\rm vs}$, in the form of a vortex-liquid state.
In ref.~\onlinecite{Yu2014Diamagnetic}, the anomaly in $\kappa_{\rm xx}$ was interpreted as a transition to a pair-density-wave phase.

\begin{figure}[h!]
\centering
\includegraphics[width=0.42\textwidth]{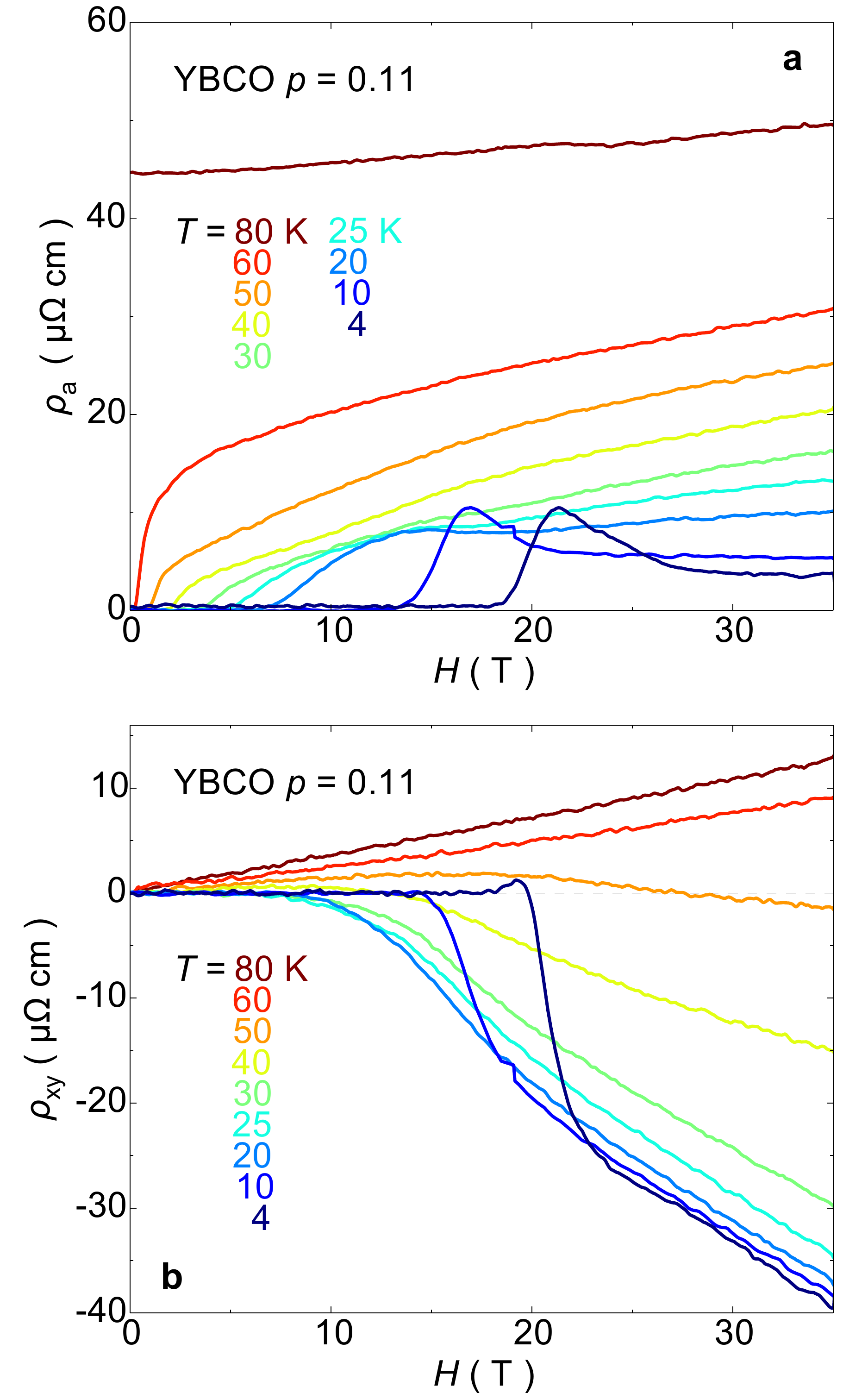}
\caption{(\textbf{a}) Longitudinal ($\rho_{\rm xx} = \rho_{\rm a}$) and (\textbf{b}) transverse (Hall; $\rho_{\rm xy}$) electrical resistivities of our $a$-axis YBCO sample with $p = 0.11$ ($y = 6.54$), plotted as a function of field up to $H = 35$~T, well above the upper critical field $H_{\rm c2}(0) = 24$~T, at various temperatures as indicated. These data are used to obtain the electrical Hall coefficient $R_{\rm H}$ and conductivity $\sigma_{\rm xy}$ in Fig. \ref{fig_2}.}
\label{fig_S1}
\end{figure}

To help resolve this debate, and more generally shed light on the nature of the normal state, we have turned to the Wiedemann-Franz law, a fundamental law of electrons in metals. It states that the conduction of heat and charge are equal in the limit of $T = 0$, where all scattering is elastic, so that $\kappa / T = L_{\rm 0} \sigma$, where $\kappa$ is the thermal conductivity tensor, $\sigma$ the electrical conductivity tensor, and $L_{\rm 0} \equiv (\pi^2 / 3) (k_{\rm B} / e)^2$ the Lorenz number.
(For a summary of prior tests of the Wiedemann-Franz law in cuprates, see Appendix~A.)
Superconductivity maximally violates the Wiedemann-Franz law since Cooper pairs conduct electricity perfectly but carry no entropy. In the vortex-liquid phase, the violation is no longer infinite but still present, since $\sigma$ and $\kappa$ are respectively enhanced and suppressed relative to their normal-state values. We can therefore use the law as a test for the existence of superconductivity in YBCO above $H_{\rm vs}$.

\section{METHODS}

We use the transverse (Hall) conductivities, $\kappa_{\rm xy}$ and $\sigma_{\rm xy}$, rather than the longitudinal conductivities, $\kappa_{\rm xx}$ and $\sigma_{\rm xx}$, because $\kappa_{\rm xy}$ is purely electronic and hence free of the large and ill-known phonon contribution that dominates $\kappa_{\rm xx}$. In the Hall channel, the law is given by $\kappa_{\rm xy} / T = L_{\rm 0} \sigma_{\rm xy}$, with $\sigma_{\rm xy} = \rho_{\rm xy} / (\rho_{\rm xx} \rho_{\rm yy} + \rho_{\rm xy}^2)$, where $\rho_{\rm xx} = \rho_{\rm a}$ and $\rho_{\rm yy} = \rho_{\rm b}$ are the longitudinal resistivities along the $a$ and $b$ axes of the orthorhombic structure, respectively, and $\rho_{\rm xy}$ is the transverse (Hall) resistivity.

\textbf{Samples}.
Our comparative study of heat and charge transport in YBCO was performed by measuring the electrical Hall conductivity $\sigma_{\rm xy}$
and the thermal Hall conductivity $\kappa_{\rm xy}$ on the same sample, using the same contacts.
This sample was a detwinned single crystal of \ybco~with oxygen content $y = 6.54$ and a high degree of ortho-II oxygen order,\cite{liang_growth_2012}
yielding large quantum oscillations, proof of a long electronic mean free path at low temperature.
The hole concentration (doping) $p$ is obtained from the superconducting $T_{\rm c}$,\cite{liang_evaluation_2006}
defined as the temperature where the electrical resistance goes to zero.
Our sample has $T_{\rm c} = 61$~K, giving $p = 0.11$. At this particular doping, the upper critical field is at a local minimum with $H_{\rm c2} = 25$~T, making it ideal for testing the Wiedemann-Franz law since available fields of $28$ to $35$~T are sufficient to access the normal state at $T = 0$.
The sample is in the shape of a rectangular platelet, with a width $w = 0.6 $~mm (along the $b$~axis) and a thickness $t = 0.1 $~mm (along the $c$~axis).
Six contacts were applied in the standard geometry, using diffused gold pads.
The current (electrical or thermal) was made to flow along the $a$~axis of the orthorhombic crystal structure,
using contacts that covered the ends of the sample to ensure uniformity.
The longitudinal electrical resistivity $\rho_{\rm xx} = \rho_{\rm a}$ and the longitudinal thermal gradient $dT_{\rm x}$ were both measured
using the same two contacts on one side of the sample, separated by a distance $L = 0.8$~mm (along the $a$~axis).
The transverse electrical resistivity $\rho_{\rm xy}$ and the transverse thermal gradient $dT_{\rm y}$ were both measured
using the same two contacts on opposite sides of the sample, separated by a distance $w = 0.6$~mm (along the $b$~axis).


\begin{figure}[h!]
\centering
\includegraphics[width=0.42\textwidth]{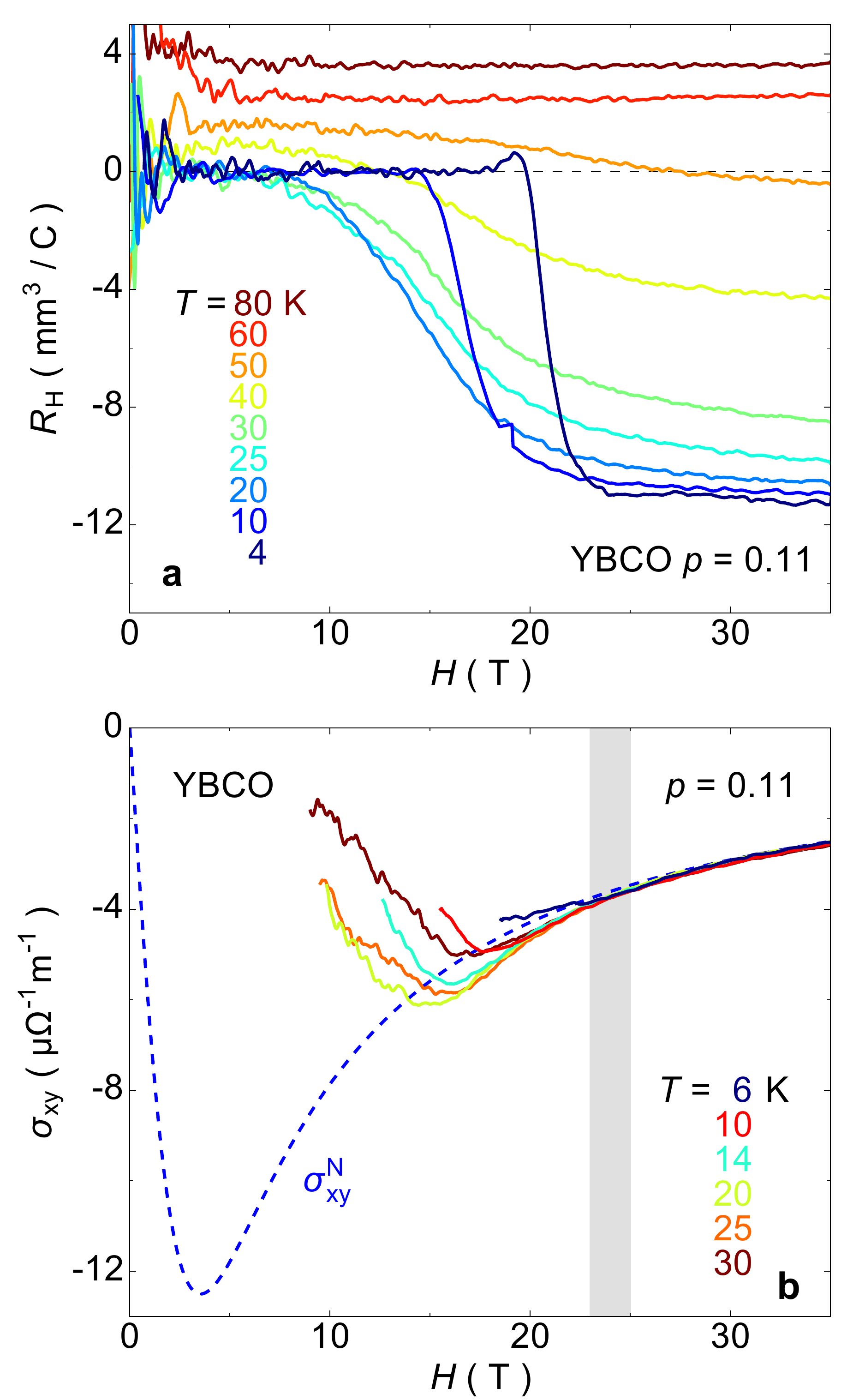}
\caption{
(\textbf{a})
Electrical Hall coefficient $R_{\rm H}$ as a function of magnetic field $H$, measured on the same sample of YBCO ($p = 0.11$)
on which $\kappa_{\rm xy}$ was measured, at various temperatures as indicated.
The Fermi-surface reconstruction causes $R_{\rm H}$ at high field to go from a positive value at $T = 80$~K to a large negative value at $T = 4$~K,
evidence that a small high-mobility electron pocket emerges upon cooling.\cite{Leyraud2007Quantuma,Leboeuf2007Electron,taillefer_fermi_2009,Leboeuf2011Lifshitz}
Note that $R_{\rm H}(H)$ at $T = 4$~K is constant above $H_{\rm c2}(0) = 24$~T, showing that there is no flux flow due to vortices above that field.
(\textbf{b})
Hall conductivity $\sigma_{\rm xy}$~vs~$H$, for temperatures as indicated, obtained from measurements of $\rho_{\rm xx}$ and $\rho_{\rm xy}$~(Fig.~\ref{fig_S1}).
Note that above $H_{\rm c2} = 24 \pm 1$~T (vertical grey band) the various isotherms of $\sigma_{\rm xy}$ collapse onto the same curve.
The dashed blue line is a plot of the normal-state conductivity $\sigma_{\rm xy}^{\rm N} = \rho_{\rm xy} / (\rho_{\rm xx}^2 + \rho_{\rm xy}^2)$,
where $\rho_{\rm xy} = R_{\rm H} H$, with $R_{\rm H} = - 11.5$ mm$^3$~/~C, and $\rho_{\rm xx} = 5$~$\mu\Omega$~cm,
values appropriate for the high-field normal state at $T = 10$~K (Fig. \ref{fig_2}a and Fig. \ref{fig_S1}).}
\label{fig_2}
\end{figure}

\textbf{Electrical transport coefficients}.
The transverse Hall conductivity $\sigma_{\rm xy}$ of our orthorhombic crystal is given by
$\sigma_{\rm xy} = \rho_{\rm xy} / (\rho_{\rm xx} \rho_{\rm yy} + \rho_{\rm xy} \rho_{\rm yx})$,
where $\rho_{\rm xx}$ and $\rho_{\rm yy}$ are the longitudinal resistivities along the $x$ and $y$ directions,
{\it i.e.}~the $a$ and $b$~axis, and $\rho_{\rm xy}$ and $\rho_{\rm yx}$ are the transverse resistivities.
We take the latter to be equal, namely $\rho_{\rm xy} = \rho_{\rm yx}$, or $\sigma_{\rm xy} = \sigma_{\rm yx}$,
consistent with $\kappa_{\rm ab} = \kappa_{\rm ba}$ (see Appendix~B).
We also assume that $\rho_{\rm yy} = \rho_{\rm xx}$, {\it i.e.}~$\rho_{\rm b }= \rho_{\rm a }$,
as observed just above $T_{\rm c}$ in similar YBCO crystals.\cite{Leyraud2007Quantuma}
The latter assumption has no impact on our test of the Wiedemann-Franz law, since at high $H$ and low $T$ we observe that $\rho_{\rm xy}^2 \gg \rho_{\rm xx}^2 \sim{~} \rho_{\rm xx} \rho_{\rm yy}$ (Fig. \ref{fig_S1}).

The coefficients $\rho_{\rm xx}~(= \rho_{\rm a})$ and $\rho_{\rm xy}$~were measured in magnetic fields up to 35~T at the NHMFL in Tallahassee,
using an AC 4-terminal method and applying the usual symmetrisation ($R_{\rm xx} = [V_{\rm x} (+H) + V_{\rm x} (-H)] / 2I_{\rm x}$)
and anti-symmetrisation ($R_{\rm xy} = [V_{\rm y} (+H) - V_{\rm y} (-H)] / 2I_{\rm x}$) procedures with respect to field direction.
The electrical current ($I_{\rm x}$) was applied along the $a$~axis.
The resulting data are displayed in Fig. \ref{fig_S1}.

\textbf{Thermal transport coefficients}.
The thermal Hall conductivity $\kappa_{\rm xy}$ of our two YBCO samples was measured at the LNCMI in Grenoble up to 28~T at temperatures below 1~K and at the NHMFL in Tallahassee up to 35~T at temperatures from 4~K to 68~K.

A constant heat current $Q_{\rm x}$ was sent in the basal plane of the single crystal, generating a longitudinal temperature difference $dT_{\rm x}$ and, in a magnetic field applied along the $c$~axis, a transverse temperature difference $dT_{\rm y}$. The thermal Hall conductivity is defined as $\kappa_{\rm xy} = \kappa_{\rm yy} (dT_{\rm y} / dT_{\rm x} ) (L / w)$, where $\kappa_{\rm yy}$ is the longitudinal thermal conductivity along the $y$~axis (perpendicular to the $x$~axis). At all temperatures, we employed a one-heater-two-thermometers steady-state method to measure $dT_{\rm x}$, using Cernox sensors calibrated in situ as a function of temperature and magnetic field. Below 4~K, $dT_{\rm y}$ was measured with a calibrated Cernox sensor. At 4~K and above, $dT_{\rm y}$ was measured using a differential type-E thermocouple known to have a weak magnetic field dependence. At $T = 10$~K, data obtained using the thermocouple were compared to data obtained using a Cernox sensor, in otherwise identical conditions, and the agreement was excellent.
%
For further details, see Appendix~C.

\textbf{Error bars}.
The error on the magnitude of $\sigma_{\rm xy}$ comes from the uncertainty in determining the geometric factor associated with sample dimensions and contact separation, estimated to be $\pm~10$ \%.
The error on the magnitude of $\kappa_{\rm xy}$ includes a similar uncertainty on the geometric factor, to which is added an uncertainty of $\pm~10$ \%
associated with thermometry, for a total of $\pm~20$ \%.

\section{RESULTS}

In Fig. \ref{fig_2}a, we show the Hall coefficient $R_{\rm H} = \rho_{\rm xy} / H$ measured in a single crystal of YBCO with a doping $p~=~0.11$
($T_{\rm c} = 61$~K), plotted as a function of field $H$ up to 35~T, at different temperatures $T$.
Note that $R_{\rm H}(H)$ at $T = 4$~K is flat above $H = 24$~T. This is a first strong evidence that there is no flux flow, and hence no long-lived vortices
above $H_{\rm c2} = 24$~T.
In Fig. \ref{fig_2}b, we see that the low-$T$ isotherms of $\sigma_{\rm xy}(H)$ collapse onto a single curve for $H >$~24~T,
given by $\sigma_{\rm xy} = 1 / \rho_{\rm xy} = 1 / (R_{\rm H} H)$, with $R_{\rm H} = -11.5$~mm$^3$~/~C (dashed blue line),
the value of the Hall coefficient at high $H$ and low $T$ (Fig. \ref{fig_2}a).
%
%
The collapse of the various isotherms at high $H$ provides a convenient way to detect superconductivity as $H$ is reduced. Indeed, superconductivity is expected to produce strongly $T$ and $H$ dependent deviations in $\sigma_{\rm xy}$, as indeed it does below $\sim 24$~T (Fig. \ref{fig_2}b).

\begin{figure}[t]
\centering
\includegraphics[width=0.42\textwidth]{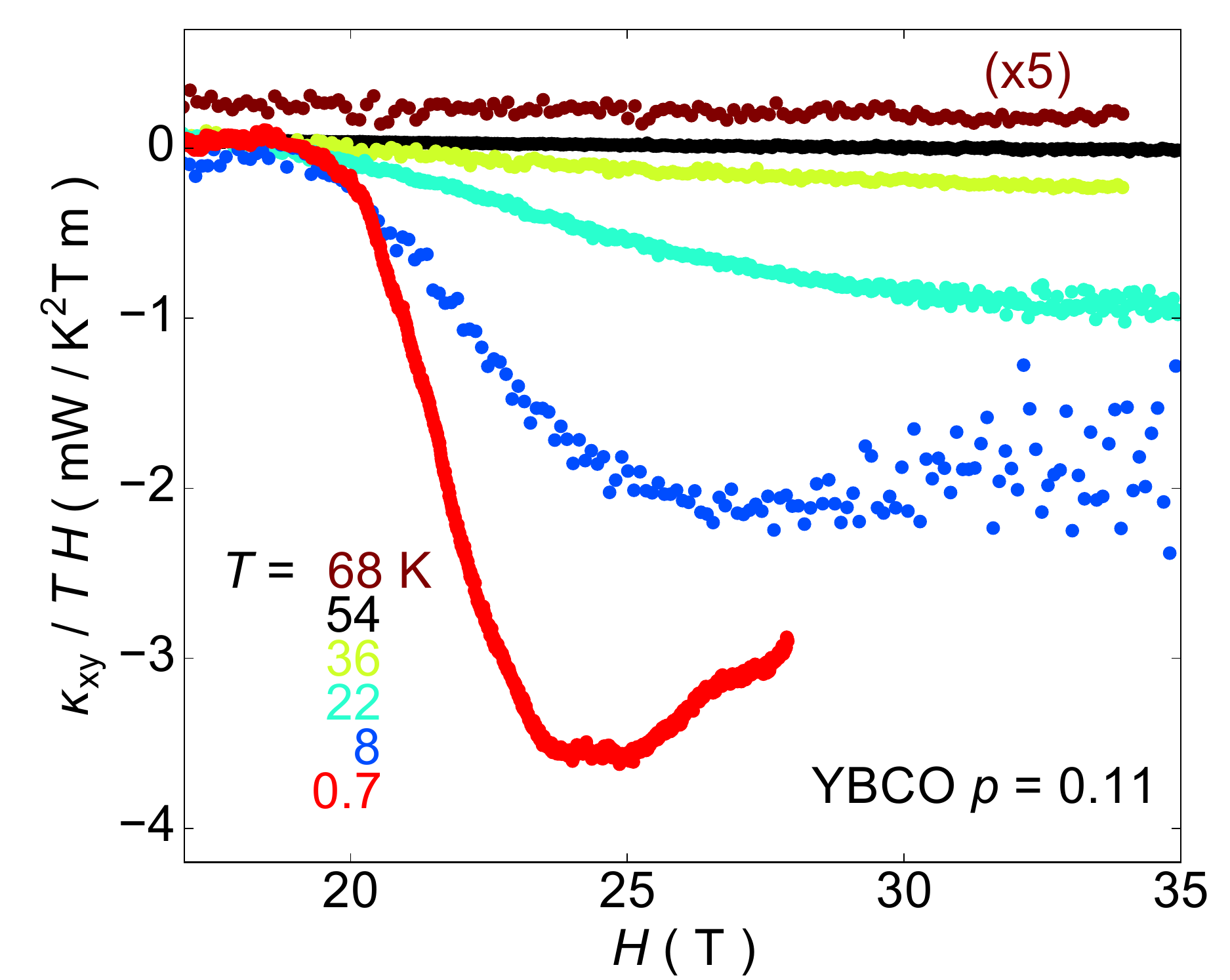}
\caption{
Thermal Hall conductivity $\kappa_{\rm xy}$ for YBCO with $p = 0.11$ ($y = 6.54$), plotted as $\kappa_{\rm xy} / (T H)$~vs~$H$
at different temperatures as indicated.
The 68~K isotherm (brown) is multiplied by a factor 5 to make it visible above the 54~K isotherm (black).
}
\label{fig_S2}
\end{figure}

\begin{figure}[t]
\centering
\includegraphics[width=0.42\textwidth]{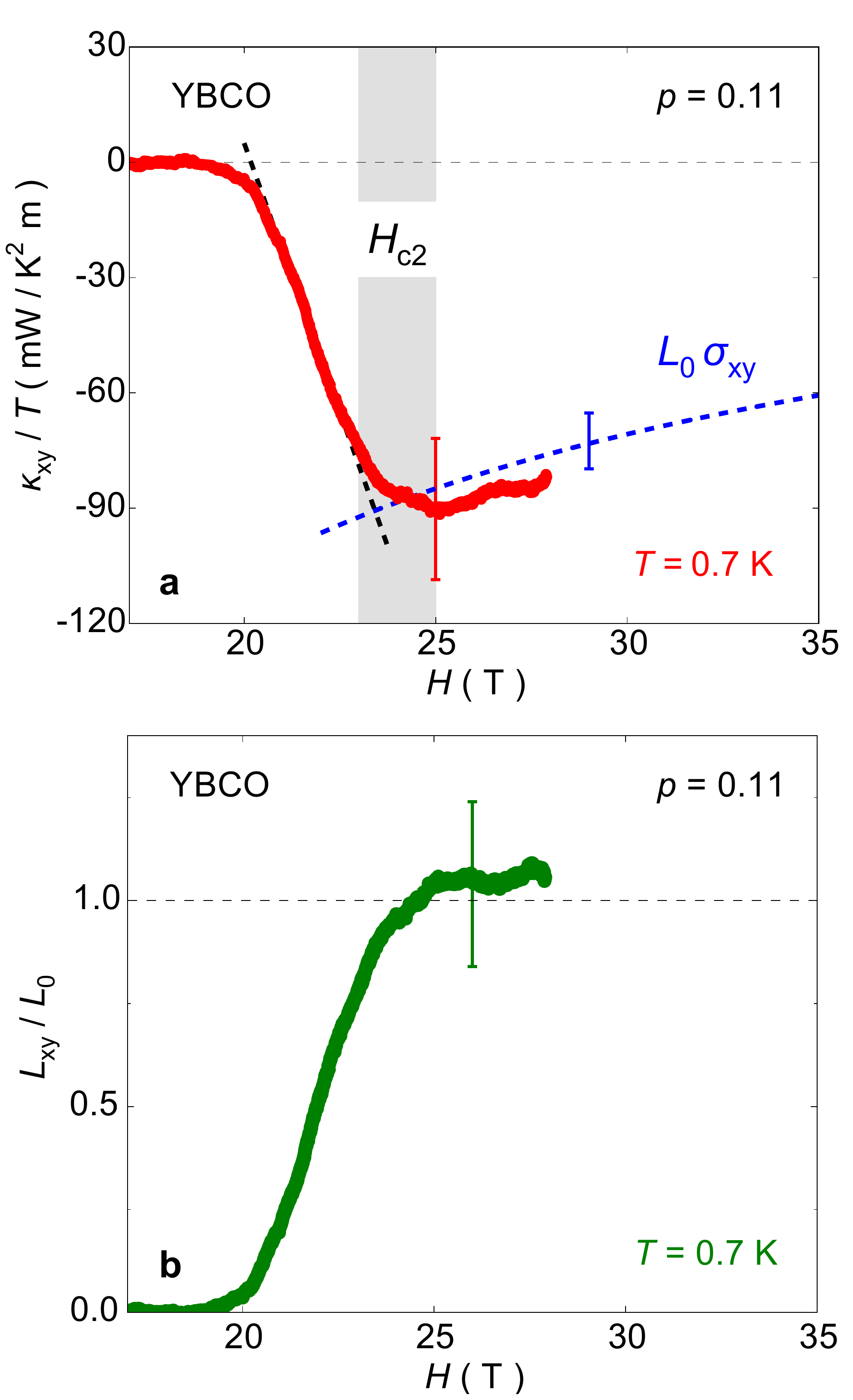}
\caption{(\textbf{a}) Thermal Hall conductivity $\kappa_{\rm xy}$ of YBCO at $p = 0.11$, plotted as $\kappa_{\rm xy} / T$~vs~$H$, at $T = 0.7$~K.
The straight dashed black line is a linear fit to the rapid rise in $|\kappa_{\rm xy}|$~vs~$H$.
A lower bound on the upper critical field $H_{\rm c2}$ is the deviation from linearity at 23~T,
while an upper bound is the minimum in $\kappa_{\rm xy}$ at 25~T, so that $H_{\rm c2} = 24 \pm 1$~T (vertical grey band).
The dashed blue line is the same as in Fig.~\ref{fig_2}b, but multiplied by the constant $L_{\rm 0} = \pi^2 / 3 (k_{\rm B} / e)^2$,
giving the measured value of $L_{\rm 0} \sigma_{\rm xy}$ at low temperature,
above $H_{\rm c2}$. The fact that $\kappa_{\rm xy} / T = L_{\rm 0} \sigma_{\rm xy}$ for $H > H_{\rm c2}$, within error bars,
shows that the Wiedemann-Franz law is satisfied -- compelling evidence that superconductivity is entirely suppressed
and the normal state is fully reached at $H_{\rm c2} = 24$~T.
Error bars on $\sigma_{\rm xy}$ and $\kappa_{\rm xy}$ are defined in Methods.
(\textbf{b})~Lorenz ratio $L_{\rm xy} = \kappa_{\rm xy} / (\sigma_{\rm xy} T)$, plotted as $L_{\rm xy} / L_{\rm 0}$~vs~$H$. The isotherm at $T = 0.7$~K is used for $\kappa_{\rm xy} / T$ (Fig. \ref{fig_3}a); the isotherm at $T = 10$~K is used for $\sigma_{\rm xy}$ (Fig. \ref{fig_2}b). $L_{\rm xy}$ saturates above $\sim 25$~T, to a value $L_{\rm xy} / L_{\rm 0} = 1.1 \pm 0.2$, showing that the Wiedemann-Franz law ($L_{\rm xy} = L_{\rm 0}$; dashed line) holds, within error bars, when $H > H_{\rm c2}$.}
\label{fig_3}
\end{figure}

The thermal Hall conductivity $\kappa_{\rm xy}$ was measured on the same single crystal on which $\sigma_{\rm xy}$ was measured (see Methods).
The various isotherms of $\kappa_{\rm xy}$ are displayed in Fig.~\ref{fig_S2}.
Looking at the lowest isotherm, at $T = 0.7$~K, we see that $\kappa_{\rm xy}$ is large and negative above 20~T, consistent with the negative electrical Hall and Seebeck coefficients,
all showing that a high-mobility electron pocket dominates the transport
properties of YBCO.\cite{Leboeuf2007Electron, taillefer_fermi_2009, Chang2010Nernst, Laliberte2011FermiSurface, Leboeuf2011Lifshitz}
With decreasing $H$, $|\kappa_{\rm xy}|$ decreases rapidly, to become negligible below 20~T or so.
We attribute this decrease to a loss of heat-carrying quasiparticles\cite{Zhang2001Giant}  and the onset of vortex scattering.\cite{Grissonnanche2014Directa}
The onset field for this decrease is $H = 24 \pm 1$~T (Fig.~\ref{fig_3}a),
in excellent agreement with prior estimates of $H_{\rm c2}$ from $\kappa_{\rm xx}$ measurements on
similar YBCO samples\cite{Grissonnanche2014Directa}~(Fig. \ref{fig_1}).

\section{DISCUSSION}

In Fig. \ref{fig_3}a, we compare the isotherm at $T = 0.7$~K, plotted as $\kappa_{\rm xy} / T$ vs $H$, with its electrical counterpart,
plotted as $L_{\rm 0} \sigma_{\rm xy}$ vs $H$.
Here $\sigma_{\rm xy}$ is simply the common normal-state curve observed at low temperature (Fig.~\ref{fig_2}b).
In Fig. \ref{fig_3}b, we plot the ratio of the two, namely the normalized Lorenz ratio $L_{\rm xy} / L_0$.
We see that the Wiedemann-Franz law is satisfied
for $H > H_{\rm c2}$, within error bars.
(Note that the law was only tested at $p = 0.11$, and therefore, strictly speaking, it is only established for the field-induced CDW state.
\cite{gerber_three-dimensional_2015,Chang2015Magnetic,Grissonnanche2015Onset})
This has two important implications for the normal state of underdoped cuprates.
First, it shows that quasiparticles conduct heat and charge just as they do in a normal Fermi liquid.
This is consistent with other signatures of Fermi-liquid behaviour in YBCO,
such as the temperature dependence of quantum oscillations\cite{sebastian_fermi-liquid_2010}
and the $T^2$ electrical resistivity at low temperature.\cite{Leboeuf2011Lifshitz,rullier-albenque_total_2007}
In general, it puts a clear and robust constraint on the nature of the low-energy excitations in the pseudogap phase of underdoped cuprates.

Secondly, it excludes the possibility of a vortex liquid above $H_{\rm vs}$ at $T \rightarrow 0$.
This means that the interpretation of the specific heat\cite{Riggs2011Heat} and magnetization\cite{Yu2014Diamagnetic} of underdoped YBCO must be re-examined. In fact, recent specific heat data \cite{marcenat_calorimetric_2015} now suggest a saturation at high magnetic fields, consistent with having no significant superconducting contribution, as our transport data show.
The effect of FSR on the normal-state susceptibility should be considered, especially as the observed drop in magnetization with decreasing temperature\cite{Yu2014Diamagnetic} occurs in tandem with the growth in CDW modulations.\cite{hucker_competing_2014,blanco-canosa_resonant_2014}
The fact that the Wiedemann-Franz law is obeyed in underdoped YBCO places strict limits on various proposed pair-density-wave states,
in which pairing coexists with CDW modulations. \cite{Chen2004Pair,berg_dynamical_2007,Lee2014Amperean,agterberg_checkerboard_2015}
Note that even though the vortex state ends at $H_{\rm c2}$, superconducting fluctuations can exist beyond $H_{\rm c2}$.\cite{Chang2012Decrease,Tafti2014Nernst}
In YBCO at $p =$~0.11 - 0.12, they are detected up to $\sim 30$~T in the low-temperature magnetization\cite{Yu2014Diamagnetic,yu_magnetization_2015} and Nernst signal.\cite{Chang2010Nernst}
However, these fluctuations appear to make no detectable contribution to $\kappa_{\rm xx}$, $\kappa_{\rm xy}$, $\rho_{\rm xx}$ or $\rho_{\rm xy}$.

\begin{figure}[t]
\centering
\includegraphics[width=0.4\textwidth]{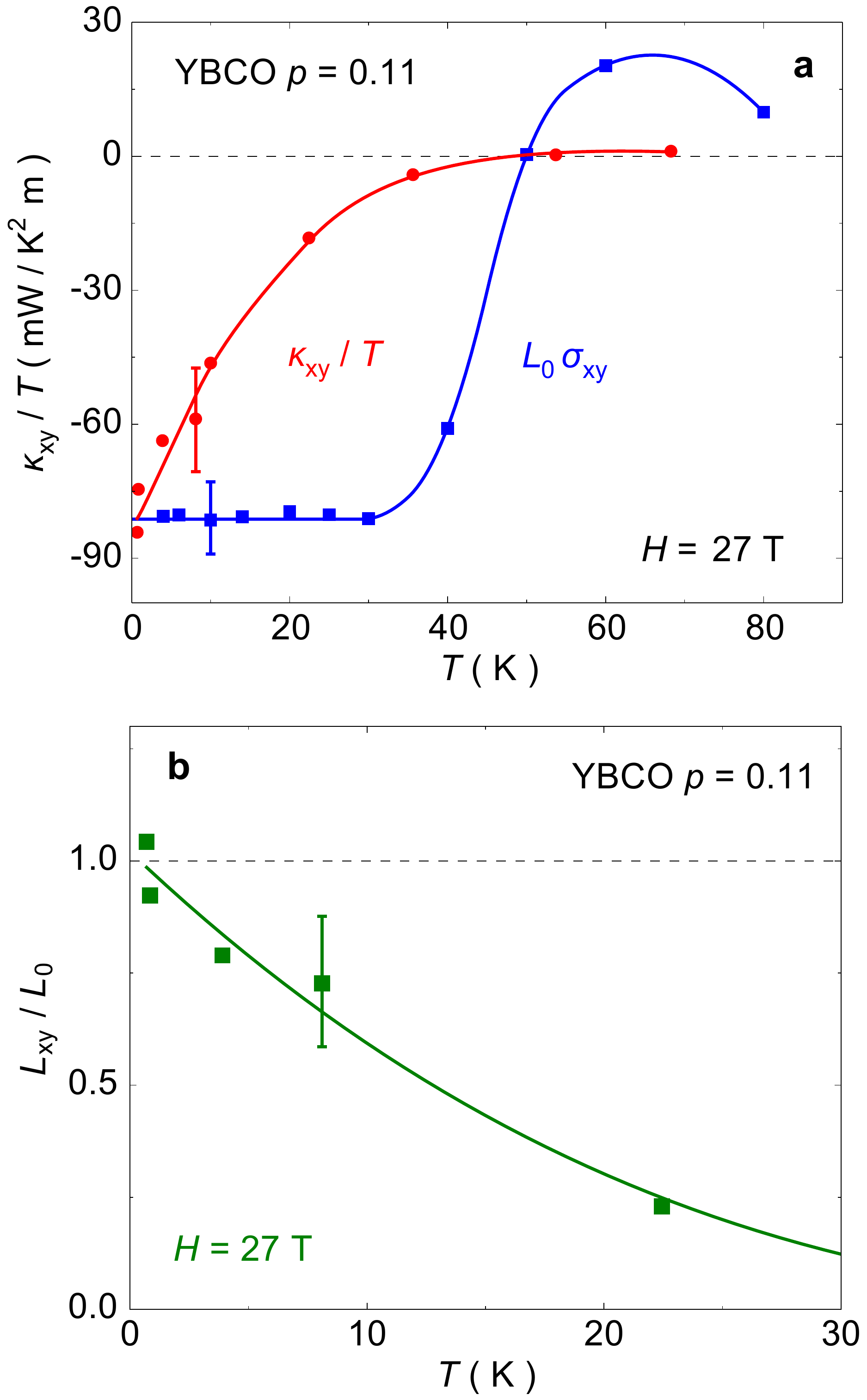}
\caption{
(\textbf{a})
Electrical and thermal Hall conductivities of YBCO as a function of temperature, measured on the same sample with  $p = 0.11$,
plotted as $L_{\rm 0} \sigma_{\rm xy}$ (blue squares) and $\kappa_{\rm xy} / T$ (red circles), respectively,
for $H = 27$~T~$ > H_{\rm c2}(0)$.
The $\sigma_{\rm xy}$ data are obtained from isotherms of $\rho_{\rm xx}$ and $\rho_{\rm xy}$ (Fig. \ref{fig_S1});
the $\kappa_{\rm xy}$ data are obtained from isotherms in Figs. \ref{fig_S2} and \ref{fig_S4}.
(\textbf{b})~Lorenz ratio $L_{\rm xy} = \kappa_{\rm xy} / (\sigma_{\rm xy} T)$, plotted as $L_{\rm xy} / L_{\rm 0}$~vs~$T$ (green squares).
The Wiedemann-Franz law ($L_{\rm xy} = L_{\rm 0}$; dashed line) is seen to hold in the limit of $T = 0$ where scattering is elastic.
With increasing $T$, however, $|\kappa_{\rm xy} / T|$ decreases rapidly even though $|L_{\rm 0} \sigma_{\rm xy}|$ remains constant,
at least initially, reflecting the effect of inelastic scattering in the normal state of underdoped YBCO.
%
}
\label{fig_4}
\end{figure}

Having established that $\kappa_{\rm xy} / T = L_{\rm 0} \sigma_{\rm xy}$ in the $T = 0$ limit,
we now examine how $\kappa_{\rm xy} / T$ and $L_{\rm 0} \sigma_{\rm xy}$ separately evolve with increasing $T$,
as a result of inelastic scattering.
In Fig. \ref{fig_4}a, we plot $\kappa_{\rm xy} / T$ and $L_{\rm 0} \sigma_{\rm xy}$~vs~$T$ at $H =$~27~T, above $H_{\rm c2}$.
As noted earlier, $L_{\rm 0} \sigma_{\rm xy}$ remains constant up to 30~K.
In sharp contrast, over the same $T$ interval, $\kappa_{\rm xy} / T$ decreases in magnitude by a factor~10.
An electrical current is more effectively degraded by a large momentum transfer, while a heat current can also be diminished by an energy loss
at small momentum transfer $\bm{q}$.
Consequently, the combination of a constant $L_{\rm 0} \sigma_{\rm xy}$ and a rapidly decreasing  $|\kappa_{\rm xy} / T|$ between $T = 0$ and 30~K
is an indication that the dominant inelastic scattering involves small-$\bm{q}$ processes.
We speculate that a possible candidate for a small $\bm{q}$-vector in the reconstructed Fermi surface of YBCO
is one that connects the tip of the square-shaped electron pocket and the tip of the hole-like ellipse
where the two nearly touch,\cite{doiron-leyraud_evidence_2015,Allais2014Connectinga},
at the CDW hot spot.
Inelastic scattering at this small $\bm{q}$-vector would affect precisely those regions of the Fermi surface that are responsible for the large negative Hall signal (Fig. \ref{fig_2}a), namely the tips of the electron pocket.
This process is therefore expected to rapidly make the Hall signal less negative, as observed in the thermal channel.

Another potential scenario for small-$\bm{q}$ inelastic scattering at low $T$ is fluctuations near a nematic quantum critical point.\cite{lederer_enhancement_2015}

\section{SUMMARY}

We have measured the thermal and electrical Hall conductivities of underdoped YBCO down to low temperature.
We find that the Wiedemann-Franz law is satisfied in the $T = 0$ limit.
This rules out a vortex liquid above the vortex-solid melting field $H_{\rm vs}$ at $T \rightarrow 0$.
More generally, it implies that any theory of underdoped cuprates must satisfy the Wiedemann-Franz law, a clear and robust constraint.


\begin{acknowledgments}
We thank K.~Behnia, A.~Carrington, S.~A.~Hartnoll, N.~E.~Hussey, P.~A.~Lee, C.~Proust, B.~J.~Ramshaw, A.-M.~S.~Tremblay and O.~Vafek for stimulating discussions.
L.T. thanks ESPCI-ParisTech, \'Ecole Polytechnique, Universit\'e Paris-Sud, CEA-Saclay and the Coll\`ege de France for their hospitality and support,
and the LABEX PALM for its support (ANR-10-LABX-0039-PALM), while this article was written.
Part of this work was performed at the Laboratoire National des Champs Magn\'etiques Intenses, which is supported by the French ANR SUPERFIELD,
the EMFL, and the LABEX NEXT.
Part of this work was performed at the National High Magnetic Field Laboratory, which is supported by the National Science Foundation Cooperative Agreement No. DMR-1157490, the State of Florida, and the U.S. Department of Energy.
R.L., D.A.B. and W.N.H. acknowledge support from NSERC.
L.T. acknowledges support from the Canadian Institute for Advanced Research and funding from NSERC, FRQNT, the Canada Foundation for Innovation,
and a Canada Research Chair.

\end{acknowledgments}


\appendix


\section{Prior tests of the Wiedemann-Franz law in cuprates}

The Wiedemann-Franz law, $\kappa / T = L_{\rm 0} \sigma$ at $T~=~0$, was investigated for cuprates in six prior studies:
in optimally-doped \pcco~(PCCO),\cite{hill_breakdown_2001}
in overdoped \tltwotwoone,\cite{proust_heat_2002}
in overdoped LSCO,\cite{nakamae_electronic_2003,sun_deviation_2009}
and in overdoped,\cite{bel_test_2004}
optimally-doped and underdoped Bi-2201.\cite{proust_heat_2005}
In all cases, the test was done on longitudinal conductivities ($\kappa_{\rm xx}$ and $\sigma_{\rm xx}$).
Because of the large phonon term in $\kappa_{\rm xx}$, extracting the electronic term is done by extrapolating $\kappa_{\rm xx} / T$ to $T = 0$.
In the study on PCCO, this procedure failed because of electron-phonon decoupling.\cite{smith_low-t_2005}
For all overdoped samples, the Wiedemann-Franz law was found to be valid in the field induced normal state, to within a few percent.
In the only prior study on an underdoped cuprate (Bi-2201), the Lorenz ratio was found to be larger than expected:
$L / L_{\rm 0} > 1.0$.\cite{proust_heat_2005}
However, values of $L / L_{\rm 0}$ exceeding 1.0 were observed only in samples whose normal state resistivity $\rho_{\rm a}(T)$ showed an upturn at low $T$,
achieving residual values $\rho_{\rm 0} > 200$~$\mu\Omega$~cm.\cite{proust_heat_2005}
The violation was attributed to a metal-insulator transition.
Our YBCO samples are in a completely different regime, with fully metallic behavior and $\rho_{\rm 0} = 4$~$\mu\Omega$~cm at $H > H_{\rm c2}$ (Fig.~\ref{fig_S1}).


\section{Onsager relation}

\begin{figure}[h!]
\centering
\includegraphics[width=0.42\textwidth]{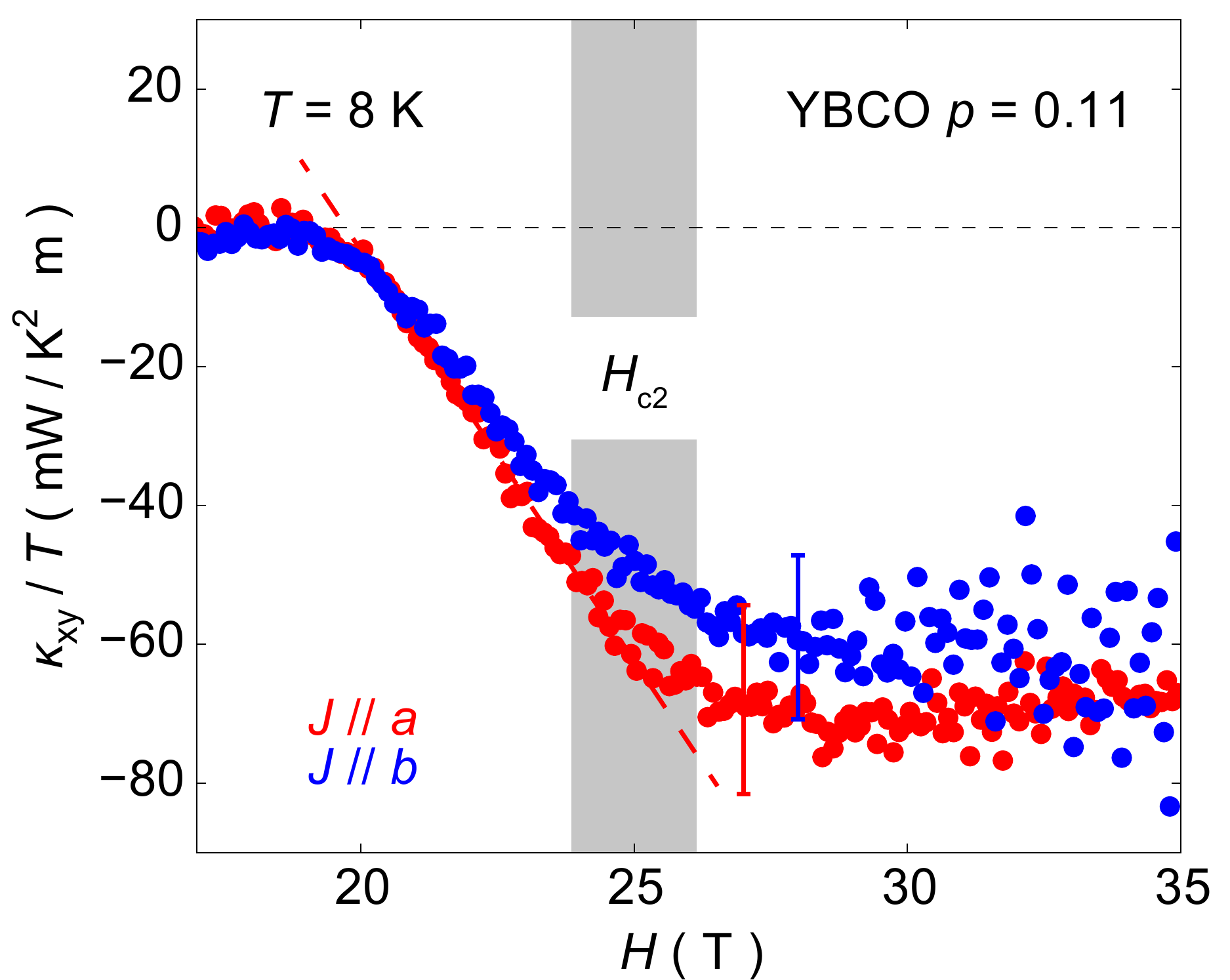}
\caption{
Thermal Hall conductivity of our two samples of YBCO with $p = 0.11$ ($y = 6.54$), at $T = 8$~K, plotted as $\kappa_{\rm xy} / T$~vs~$H$.
In one sample, the heat current flows along the $a$~axis of the orthorhombic crystal structure (red, $J \parallel a$),
while in the second it flows along the $b$~axis (blue, $J \parallel b$) (Methods).
Within error bars ($\pm~20$~\%), we see that both samples yield the same curve, so that $\kappa_{\rm ab} = \kappa_{\rm ba}$, as expected from the Onsager reciprocity relation. %
The dashed line is a linear fit to the steepest part of $|\kappa_{\rm xy}|$~vs~$H$.
We define $H_{\rm c2}$ as the field above which $|\kappa_{\rm xy}|$ departs from that linear rise (vertical grey band), giving $H_{\rm c2} = 25 \pm 1$~T,
as plotted in Fig.~1.
}
\label{fig_S3}
\end{figure}

Because YBCO has an orthorhombic crystal structure, measurements on two samples are necessary to obtain $\kappa_{xy}$:
one with a current along $x = a$ and one with a current along $x = b$. For oxygen content $y = 6.54$, two nominally identical samples were used, with their length along the $a$~axis and the $b$~axis, respectively. So $\kappa_{\rm ab}$ was measured on the first sample, using $\kappa_{\rm bb}$ measured on the second, and $\kappa_{\rm ba}$ was measured on the second sample, using $\kappa_{\rm aa}$ measured on the first. Within error bars, we find that $\kappa_{\rm ab} = \kappa_{\rm ba}$ at all fields and temperatures, thereby satisfying the Onsager relation, as shown in Fig. \ref{fig_S3}.

\begin{figure}[h!]
\centering
\includegraphics[width=0.42\textwidth]{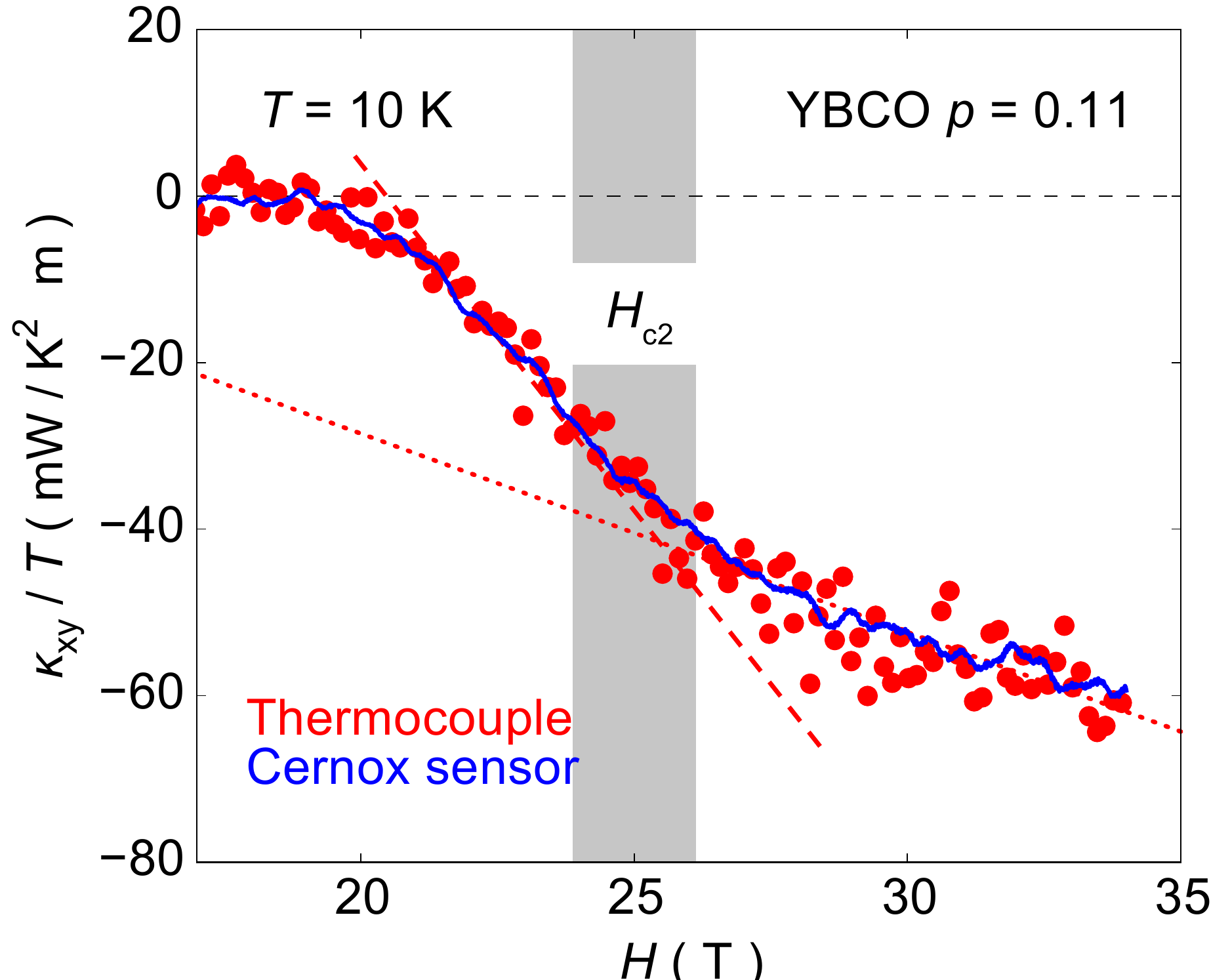}
\caption{
Thermal Hall conductivity $\kappa_{\rm xy}$ as a function of field at $T = 10$~K obtained from two different measurements
of the transverse temperature difference $dT_{\rm y}$:
using 1) a Cernox sensor (blue curve) and
2) a type-E thermocouple (red dots).
In both cases, the longitudinal temperature difference $dT_{\rm x}$ is measured with two Cernox sensors.
The two measurements of $\kappa_{\rm xy}$ show excellent agreement, confirming that our thermometry in high fields is reliable.
The dashed line is a linear fit to the steepest part of $|\kappa_{\rm xy}|$ vs $H$. We define as $H_{\rm c2}$ the field above which $|\kappa_{\rm xy}|$ departs from that linear rise (vertical grey band), giving $H_{\rm c2} = 25 \pm 1$~T (Fig.~1).
The dotted line shows the linear behaviour ($\kappa_{\rm xy} \sim H$) expected of a metal when $\kappa_{\rm xy}$ becomes comparable to or smaller than $\kappa_{\rm xx}$.}
\label{fig_S4}
\end{figure}


\section{Thermometry}

The longitudinal temperature difference $dT_{\rm x} = T_{\rm hot} - T_{\rm cold}$ was measured using Cernox resistive sensors positioned on one side of the sample near the hot ($T_{\rm hot}$) and cold ($T_{\rm cold}$) ends. In zero field, the sensors are calibrated in-situ against a reference calibrated Cernox sensor. At $T < 15$~K or so, Cernox sensors show a pronounced (negative) magnetoresistance. In order to properly determine $T_{\rm hot}$ and $T_{\rm cold}$ in a finite field, the hot and cold Cernox sensors were calibrated by performing field sweeps at different closely spaced temperatures between $0.5$~K and $15$~K. The probe temperature was kept constant when sweeping the magnetic field by using a strain gauge with a field-independent resistance as the temperature regulator of our probe. That the temperature was indeed kept constant was checked against a Cernox sensor independently calibrated in magnetic fields up to $27$~T and down to $1.5$~K, and also against a RuOx sensor known to have a weak and linear magnetoresistance. Below $1.0$~K, the probe temperature was kept constant against the vapour pressure of a helium-3 bath. Field sweeps going up or down gave identical traces. Above 4~K, the transverse temperature difference $dT_{\rm y}$ was measured with a type-E constantan-chromel-constantan differential thermocouple known to have a weak field dependence. Below 4~K, $dT_{\rm y}$ was measured using Cernox sensors calibrated as for the $dT_{\rm x}$ measurement. In a field $H$, $T_{\rm hot}$ contains a contribution from the transverse gradient $dT_{\rm y}$:
$T_{\rm hot} (\pm H) = T_{\rm hot}(SYM) \pm dT_{\rm y} / 2$.
By antisymetrising $T_{\rm hot}$, we get the transverse thermal gradient $dT_{\rm y}$ with a single sensor measurement.
Quantitative agreement between the two methods used to measure $dT_{\rm y}$ is demonstrated in Fig.~\ref{fig_S4}.
The excellent agreement demonstrates that our in-field thermometry is accurate and reliable.
Data are systematically taken at positive and negative fields, and $dT_{\rm x}$ and $dT_{\rm y}$ are associated
with the symmetric and anti-symmetric traces, respectively.
The magnetic field was swept at a rate of 1~T~/~min, well below the level at which thermal hysteretic effects are observed.

\end{document}